\documentclass[a4paper,10pt]{article}
\usepackage{amssymb}
\usepackage{amsmath}

\usepackage{colortbl}

\usepackage{pifont}

\usepackage{parskip} %% <-- added

\usepackage{graphicx}
\graphicspath{{pic/}}
\usepackage{setspace}

\usepackage{float}
\providecommand{\keywords}[1]{\textbf{Keywords:} #1}

\usepackage{hyperref}
\usepackage{url}

\usepackage{cleveref}

\usepackage{algorithm}
\usepackage[noend]{algpseudocode}
\makeatletter
\def\BState{\State\hskip-\ALG@thistlm}
\makeatother

\usepackage[numbers,sort]{natbib}
\usepackage{usebib}
\newbibfield{author}
\bibpunct{[}{]}{,}{n}{}{,~}
\bibinput{ref}

\usepackage{multirow}
\usepackage[table,xcdraw]{xcolor}
\usepackage{lscape}
\usepackage{booktabs}
\usepackage{hhline}

\usepackage{varwidth}
\usepackage{fancyvrb}

\usepackage{tikz}

\usepackage[utf8]{inputenc}
\usepackage{authblk}

\DeclareUnicodeCharacter{FB01}{\,}

\makeatletter
\newcommand\notsotiny{\@setfontsize\notsotiny{6.31415}{7.1828}} 
\newcommand\notsotinynew{\@setfontsize\notsotinynew{6}{7}}
\makeatother

\newcommand{\orcid}[1]{\href{https://orcid.org/#1}{\textcolor[HTML]{A6CE39}{\aiOrcid}}}

\begin{document}
\title {\Large \textbf{Lightweight Electronic Signatures and Reliable Access Control Included in Sensor Networks to Prevent Cyber Attacks from Modifying Patient Data}}
\author{Mishall Al-Zubaidie\thanks{Corresponding author: e-mail: mishall\_zubaidie@utq.edu.iq).}}
%\author[2]{Raad A. Muhajjar}
\affil{Department of Computer Sciences, Education College for Pure Sciences, University of Thi-Qar, Thi-Qar, 64001, Iraq}
%\affil[2]{Department of Computer Science, Faculty of Computer Science and Information Technology, University of Basrah, Basrah 61004, Iraq}
\date{}

%*** Specify number of column ***
%\twocolumn % \onecolumn \twocolumn

\maketitle
\small %\Large 14 \large 12 \normalsize 10 \small 9 \footnotesize 8
\noindent{\bf Abstract-} Digital terrorism is a major cause of securing patient/healthcare providers data and information. Sensitive topics that may have an impact on a patient's health or even national security include patient health records and information on healthcare providers. Health databases and data sets have been continually breached by many, regular assaults, as well as local and remote servers equipped with wireless sensor networks (WSNs) in diverse locations. The problem was addressed by some contemporary strategies that were created to stop these assaults and guarantee the privacy of patient data and information transferred and gathered by sensors. Nevertheless, the literature analysis outlines many indications of weakness that persist in these methods. This study suggests a novel, reliable method that bolsters the information security and data gathered by sensors and kept on base station datasets. The proposed approach combines a number of security mechanisms, including symmetric cryptography for encryption, asymmetric cryptography for access control and signatures, and the Lesamnta-LW method in the signature process. Users' information is shielded from prying eyes by the careful application of these measures and a sound approach. Investigational comparisons, security studies, and thorough results show that the suggested method is better than earlier methods. 

\noindent\keywords{AES-192, ACM, data tampering, NGAC, Lesamnta-LW,  SAML, sensor networks, vampire}
\pagenumbering{arabic}

%\enlargethispage{-7mm}
\section{Introduction}
\label{sec:introduction}
\begin{figure}[t]\vspace*{4pt}
%\centerline{\includegraphics{fx1}\hspace*{5mm}\includegraphics{fx1}}
\centerline{\includegraphics[width=10cm,height=7cm]{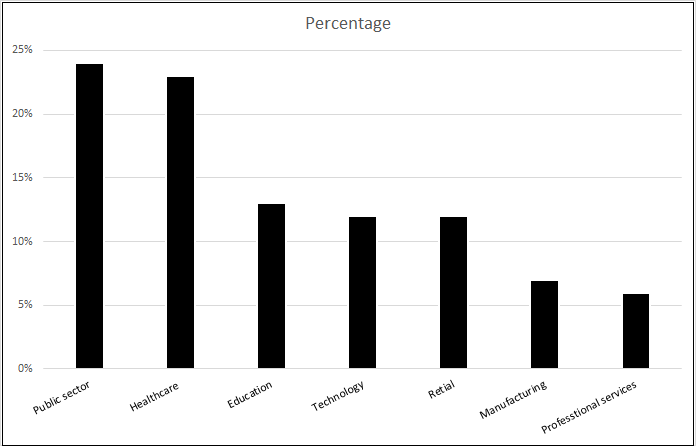}}
\caption{Cyberthreats and data breaches in several crucial industries \citep{fp58}}
\label{fig:attacks_on_sectors}
\end{figure}

Electronic applications by the Internet of Things (IoT) have provided many benefits to users and multiple organizations; One of its most important features was the expansion of services, communications, dealing with big data, and providing all kinds of information (audio, video, image, text ... etc.) for the different healthcare systems. Consequently, these systems are increasingly required to support patient-doctor communication activities that take place in that environment and directly \citep{itmts}. It is the use of patient data in digital form that increases the likelihood that some users or suspicious organizations will engage in malicious activities online that may lead to psychological, physical, moral and/or economic harm. These malicious activities became known by the term "data terrorism" or "digital terrorism". In a healthcare application scenario, these threats can be in a variety of ways, such as destroying a computer system in a health center or exposing patients' private medical data in a health database to the public. Whatever methods the attacker uses, the general and specific effects are the same: sensitive patient data and patient care are seriously compromised, the acceptability of the health care system fails, and services are not guaranteed. In addition, there is evidence indicating that cyber threats are increasing and that a large part of health care systems are not equipped to counter these threats (digital terrorism), for example, Figure~\ref{fig:attacks_on_sectors} illustrates how digital terrorism affects electronic applications across many industries, whereas the healthcare industry is subject to a higher percentage (23\%) of cyberattacks than other industries. Securing the patient data carrier is not easy because threats are constantly evolving and in a variety of ways, and with the realization that cyber terrorism must be part of a broader strategy to regulate and manage the risks of digital data, there are many mechanisms and techniques that healthcare organizations can use to protect digital health data of electronic threats \citep{adlo}.\\

The preserve privacy of medical data in the health systems such as electronic health records (EHRs) has been a major importance to academic, research and health organization, since the quality and efficiency of medical records management  \cite{fp59} by utilizing the Internet. For authorization privileges to control access to information and patients' data within EHR systems, authorization privileges are required \citep{fp60}. Accurate health information is essential for diagnosing illnesses and characterizing medical conditions when patients' medical data are transferred electronically from patient to provider \citep{fp61}. Patients suffer severe complications from any alternative to these records. Additionally, patients may experience harassment, discrimination, or even pass away if the diagnostic findings for illnesses like dermatological or HIV infection alter while being sent from sender to recipient \citep{fp62}. Terrorist actions have the potential to increase national vulnerability and insecurity through the disclosure of user records, the alteration of records, the erasure of records, or the impersonation of users inside health systems. EHR approaches should present edge-to-edge security for users' records. A central server can also enhance data management but is a tempting target for hackers because it stores medical records and authorizes users (patients and providers) by privacy policies \citep{psffl}. To prevent a breach of the server's policy and protect the privacy of patient records, privacy mechanisms are essential. 
%fig
A major challenge for authorization schemes is how to utilize patients' records for a variety of purposes like access by family members, consultations, caregivers, emergency situations and medical research  (indirect/secondary utilization). For instance, the doctor researcher must not override the permissions of privacy allowed to him/her. A patient's privacy is vigorously compromised, in case of emergency, when the physician is unavailable or the patient cannot consent to another provider  \citep{eaos}. Moreover, the patient's records must be received by a relative if the patient is incapacitated \citep{idca}. When operating on a patient, it is sometimes necessary for the doctor to consult with another physician. Medical records could be penetrated and accessed by all of these circumstances. When medical records are exchanged between patients and EHR providers, malicious threats have the potential to misuse or abuse them \citep{msop}. Numerous incidents of breaches into databases and medical records of patients have occurred, including 
\begin{itemize}
\item Attackers posing as patients gained access to SingHealth's clinic and clinic datasets on July 20, 2018, the largest health operator in Singapore. The attackers next illegitimately accessed and penetrated medical records on 4 July 2018. The medical records of 1.5 million Singaporeans inclusive of records belonging to Prime Minister Lee Hsien Loong were penetrated \citep{fp57}.
\item In May 2021, a ransomware attack breached the California hospital platform, causing their patient portal to go offline and EHR activities to be interrupted \citep{fp55}.
\item Effective attacks to healthcare applications include the May 2021 Scripps Health cyberattack and the January 2022 Kronos incident \citep{fp56}.
\end{itemize}
The security lapses clarify why robustness and excellent security are necessary for health applications. Because every patient and provider has privileged access to a server dataset, an inside danger can likewise access a medical record more readily than an external one. The suggested protocol included next-generation access control (NGAC) to offer strong security based on roles and provider/patient characteristics. Therefore, while guaranteeing the security of medical information, health sensor systems schemes must include procedures that preserve  healthcare user/sensor information, authorization policies, and patient privacy demands \citep{rwi}. 
\subsection{Threat to Authorization Approach}
An analysis of the drawbacks of many health applications requires the construction of a threat model. The threat model \citep{fp32} and \citep{tbbs} that named Dolev-Yao is utilized to check users’ authorization in proposed approach. It is a practical/formal manner of analyzing authorization protocols in actual applications. This model is efficient in testing and analyzing different threats. This proposed supposes that threats could be passive, active, external and internal. Furthermore, it assumes that the attributes server (AS) is reliable and secure in resistance to information databases breakthrough threats.  Thus, this model addresses the following attacks:
\begin{itemize}
\item Information logging at a particularly crucial location may be interrupted if the health wireless sensor is prevented from sending data or connecting to the network (node outage).
\item Gaining unauthorized access to the health network by using a Man-in-the-Middle threat to change patient records.
\item Utilizing an impersonation threat to obtain medical records in order to send a counterfeit authorization message.
\item Increasing the pace of a rogue node during the routing process in order to obtain health sensor packets before the time is due (rushing attack).
\item Consuming the power of sensor nodes and stopping the network altogether (vampire attack).
\item Sending raw data to the wrong node and selecting the longest path to traverse (neglect and greed attack). 
\item The captured data packet might be dropped by a rogue node, which is always ready to reply to the route request (packet drop attack).
\end{itemize}
\subsection{Objectives of Study}
The authorization approach aims to achieve the following tasks: 
\begin{itemize}
\item Combining features and roles models:  The suggested method integrates two authorization models (RBAC and ABAC) in the NGAC model to facilitate the management of patient data at the roles and attributes levels;
\item Adding digital signatures to security assertion markup language (SAML) requests:   SAML representations must have integrity and anonymity for the subjects' wishes and regulations;
\item Utilizing a digital signature mechanism and secrecy sharing scheme: The suggested method for granting users and sensors permission is predicated on the exchange of Shamir secrets with digital signatures (ECDSA);
\item Combining Lesamnta-LW and ECDSA: This goal makes it possible to employ lightweight ECDSA-Lesamnta-LW signatures rather than ECDSA-SHA-1, particularly when sending data gathered by WSN; 
\item Hiding information and data stored in network requests: The proposal accomplishes encryption and decryption operations by applying AES-192 cryptography with 10 rounds to prevent attackers from gaining access to the security parameters of users and the network.
\end{itemize}

%\begin{nomenclature}
%\begin{deflist}[A]
%\defitem{WSN}            \defterm{wireless sensor network}
%\defitem{ECDSA}\defterm{elliptic curve digital signature algorithm}
%\defitem{Lesamnta-LW}\defterm{lightweight hash function}
%\defitem{EL}\defterm{ECDSA-Lesamnta}
%\end{deflist}
%\end{nomenclature}

The remainder of the research is structured as follows. In Section 2, we describe the design of QDFaultInjector, our framework for fault injection testing of kernel modules, in detail. In Section 3, we design some corresponding experiments to evaluate QDFaultInjector. In addition, QDFaultInjector is compared with several existing fault injection tools and
a brief summary is made in Section 4.

\section{Overview of Current Authorization Approaches}
\label{sec:related_work}
This section highlights the shortcomings of authorization policies intended to safeguard healthcare users inside a health network, based on relevant existing standards.\\ \\
 \citet{fp35} introduced combining the ABAC model with both XML and XACML to support electronic signatures and encryption in the EHR \citep{stuh}. Their model uses partial encryption with signature XML files to provide security for cloud environments with EHR. Their model implements two phases of using XACML to provide authorization requests and using XML with signature and encryption to provide data protection \cite{utlc}. Nonetheless, there are several privacy issues associated with a cloud environment with EHR because it is possible to use the same data from multiple organizations or health institutions which can cause privacy issues. Furthermore, the use of public key cryptography with XML requests and responses can incur a significant cost to model performance. Also, some authorization requests and responses are sent explicitly, causing some authorization parameters to be exposed by attackers.\\
 \citet{fp65} proposed a decentralized scheme for sharing health data based on privacy and security features. Their scheme leverages blockchain technology and smart contracts to support privacy. They indicated that their scheme allows for the disclosure of certain pieces of data based on patient selections and privacy preferences. However, their scheme does not provide sufficient security analysis to document the protection of patients' health records. Also, the disclosure of certain parts of health data by patients is a clear breach of privacy, as not all patients have an adequate security culture. Designing a scheme based on patients' choices/selects will not be convincing for patients on the one hand and health institutions on the other.\\ 
\citet{fp68} established an architecture to secure health record sharing in an electronic medical record (EMR) and eliminate a single-point problem of failure.  Smart contracts, access control, activity detection, and revocation were the four states that their framework employed.  They stated that massive data for health databases was managed by their framework.  The Edwards-curve digital signature algorithm (EdDSA) is used to achieve the signature, and its framework makes use of elliptic curve cryptography (ECC).  The authors did not, however, specify how well their system would defend against attacks.  Additionally, using ECC encryption in healthcare institutions when working with large amounts of data might be expensive.\\
To enhance data management and privacy in EMR records, \citet{fp69} combined public and private clouds.  They created the public-key algorithm-based EMR authorization system, which safeguards patient medical records.  They asserted that their system enables integrity, backward and forward security, non-repudiation, unlinkability, anonymity, and protection against replay, man-in-the-middle, and impersonation.  However, their approach lacks the anti-traceability and anti-leakage capabilities that are critical to any trustworthy authorization system that offers strong data privacy, as well as tools to prevent DoS and dataset assaults.\\
\citet{fp66} created a way to authorize users in the EHR health system using a blockchain and patient privacy-preserving approach to access control.  They point out that strong safeguards are necessary to stop data leaks in the health information system involving private medical information.  To secure authorization requests, their approach uses file authorization contracts and common cryptographic techniques.  Nevertheless, information regarding authorization requests and policies regarding their acceptance or rejection is not provided by this means.  Additionally, file authorization contracts for source-restricted devices like WSN do not depend on a thin performance support mechanism.\\
Similar to \citep{fp68}, \citet{fp67} suggested a decentralized file healthcare system for EHR that applies a threshold signature to eliminate the problems of DoS attacks and a single point of failure. They proposed a consensus manner for the design of Istanbul Byzantine fault tolerant (IBFT). Although a decentralized system design is beneficial in reducing a single point of failure, their system can cause copies of unwanted and redundant authorization requests and the system will require larger resources to implement. Also, the authors address the danger of DoS but do not address the dangers of impersonation and internal attackers accurately.\\
According to \citet{fp64}, they suggested a privacy plan for health systems that supports scalability, dynamism, and security risk tolerance by utilizing the RBAC approach and multi-factor authentication.  Their plan employed symmetric encryption and hash methods to authenticate users and an asymmetric encryption system that relied on Schnorr's signature to offer multi-factor authentication for the administrator.  Their plan, however, clearly violated security as it only protected the administrator's requests and not those of other users.  Furthermore, their schema only used user roles to safeguard permission requests, which was insufficient to support privacy because it only used the RBAC approach.\\
Lastly, \citet{fp61} suggested creating a patient record management strategy to satisfy e-health standards.  They looked into applicability and throughput concerns.  The authors also demonstrated the applicability of their approach by comparing its security to that of the Robust Healthcare-Based Blockchain (SRHB).  They noted that privacy is crucial in e-health and cloud computing environments, and that the majority of health applications conceal health data using AES-128 encryption.  However, the authors neglected to mention that internal attacks pose a greater risk to health authorization systems than external ones.  Additionally, their plan lacked a safeguard against alteration and a mechanism for data integrity.

\section{Introductory Conceptions of Authorization Mechanisms}
\label{sec:prelininary}
To provide a reliable authorization scheme, it requires studying privacy mechanisms and clarifying some details in building user authorization rules and policies in EHR health applications. Therefore, privacy concepts require the adoption of robust mechanisms to support the privacy preservation of data collected from WSN and the security of data stored in servers. The data collected by WSN is transmitted in an insecure medium by IoT in EHR systems which also requires privacy mechanisms for preserving patient data. In this section, a set of basic mechanisms are briefly explained:
\begin{itemize}
\item Symmetric encryption mechanism\\
There are many symmetric encryption algorithms that rely on a shared key in protecting records in health databases, the most prominent of which is the advanced encryption standard (AES) or Rijndael, introduced in 2001 by the National Institute of Standards and Technology (NIST) \citep{faas}. AES encryption is an upgraded version of the data encryption standard (DES) and Triple DES, proposed in January 1999. AES was designed to eliminate the security problems that DES had with compromised attacks where DES uses a 56-bit key. The AES algorithm is a 128-bit block cipher, also, this algorithm has different key lengths to block attacks such as 128, 192, and 256. Figure~\ref{fig:aes_algorithm} shows the encryption and decryption processes in the AES algorithm. Each block cipher is split into 16 bytes in a 4*4 state array. This algorithm consists of 10, 12 and 14 rounds. To mix the data with the symmetric encryption key, each round includes Substitute Bytes, Shift Rows, Mix Columns, and Add Round Key processes \citep{fp70, bacf}.\\
Substitute Bytes process is nonlinear transformation and invertible, during this process, the 8-bit (byte) of the state array is commuted with a byte in the substitution box (S-box). This process uses a lookup table that contains 28 words divided into bytes. This step is useful to prevent the connection between the encrypted data and the encryption key from being exposed. Using 8-bit input data, S-box locations are addressed based on the most significant and least significant nibbles. In the Shift Rows process, a predefined offset is applied to each byte in the state array rows \citep{cfca}. There is no change to each of the bytes in the first row, but each of the bytes in the second row is shifted one position to the left of the original. Similarly, the third and fourth rows are shifted two and three positions to the left, respectively. Mix Columns processing is performed in all rounds except the last round. The MixColumns process, it multiplies each column by a modular polynomial in GF($2^8$), instead of calculating separately. Alternatively, the SubBytes and MixColumns processes can be joint into large Look-Up Tables (LUTs). The MixColumns process mixes the columns of the state array linearly. The Add Round Key process applies the derived round key to the data block to secure the data. After the main key has been expanded, a subkey is derived, resulting in a key array of 176 bytes in 44 words (using a key length of 128 bits(. Each byte of the block was XORed with the corresponding byte in the round-key. The decryption process in AES is performed opposite to the encryption process. Decryption takes ten rounds in AES for keys with 128 bits, twelve rounds for keys with 192 bits, and fourteen rounds for keys with 256 bits. However, the round of the decryption procedure divided into Inverse Sub Bytes, Inverse Shift Rows, Inverse Mix Columns and Add Round Key (Figure~\ref{fig:aes_algorithm} shows decryption process).
%fig
\begin{figure}[t]
	\centering
		\includegraphics[width=10cm,height=10cm]{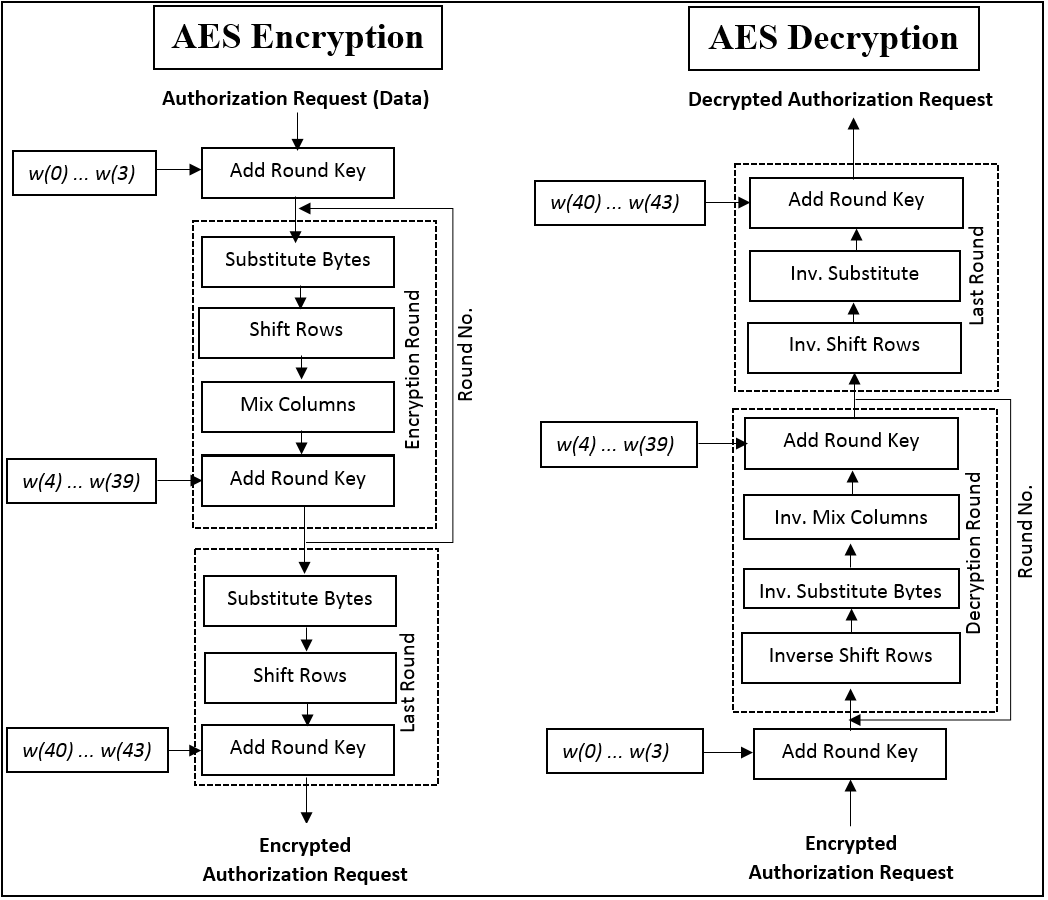}
	\caption{AES algorithm for encryption and decryption}
	\label{fig:aes_algorithm}
\end{figure}
\item Data integration mechanism\\
Integration of health records is a very important issue to ensure that patient data and reports remain unchanged \citep{etds}. ECDSA, or elliptic curve digital signature algorithm, is an asymmetric signature mechanism introduced by Scott Vanstone in 1992 \citep{rp}. This algorithm is designed to supply authentication, integrity and non-repudiation requirements.  In conventional digital signature algorithms (DSA), primes $p$ and $q$ are selected and $q|(p-1)$ and $G$ are subgroups of order $q$ of $Z * p$. Multiplication modulo $p$ is then performed on $G$. Function $H '$ is defined by $H ' (R) = R\ mod\ q$ \citep{fp72}. It contains a cyclic group of prime order $q$, a generator ($g$) for $G$, and two hash functions $H: [0, 1] /rightarrow Zq$ and $H ': G /rightarrow Zq$. ECDSA is a development of the DSA algorithm based on elliptic curves. As long as the parameters of this algorithm are carefully chosen, the ECDLP enables it to block various attacks when using points on the curve for signing data and information. This algorithm generates the private key ($Pr$) and a random number ($k$) to calculate the public key ($Pu$) through $Pu=Pr.k$. Therefore, ECDLP means that it is very difficult for an attacker to extract $k$ from $Pr$ and $Pu$. This algorithm relies on small parameters which it produces smaller keys compared to other asymmetric algorithms like RSA.This case makes this algorithm very suitable for handling big data in health institutions as well as being applicable to source-restricted devices such as WSN and RFID. Sensors and network devices require the use of a robust and high-performance mechanism to ensure data integrity because complex computations in some public key algorithms cause loss of sensor resources such as power and storage \citep{rwe,dcao}. Specifically, source-restricted devices need appropriate routing protocols \citep{rwp} and security protocols with a high level of security. Figure~\ref{fig:ecdsa_algorithm}shows signature and verification processes in ECDSA \citep{fp72}. However, the ECDSA algorithm relies mainly on SHA-1 for its operations, since SHA-1 produces a message digest of length 160 which is not very safe against attacks. In addition, SHA-1 does not offer high performance compared to lightweight hash algorithms. Therefore, one of the weaknesses of the ECDSA algorithm is SHA-1 \citep{eead}.
%fig
\begin{figure}[t]
	\centering
		\includegraphics[width=10cm,height=7cm]{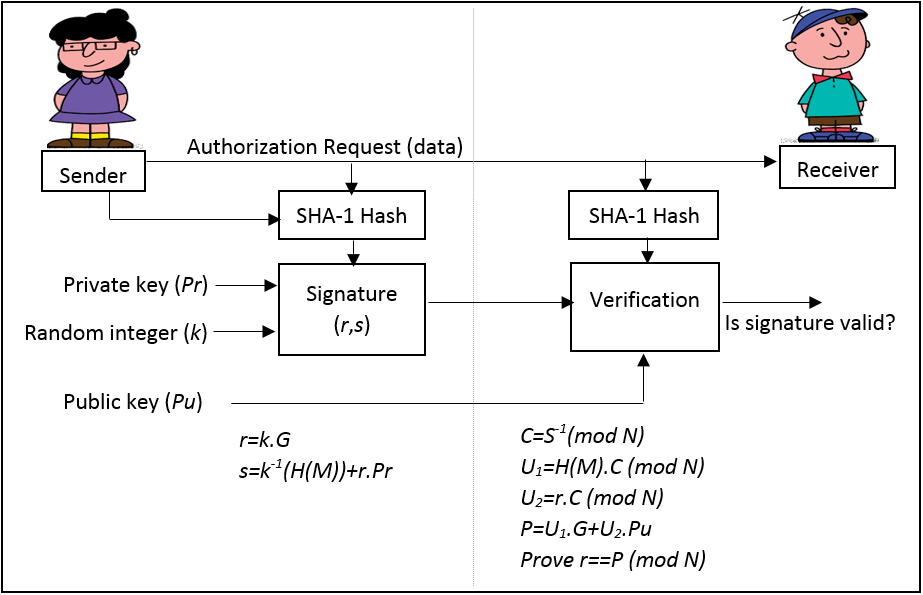}
	\caption{ECDSA algorithm for signature and verification}
	\label{fig:ecdsa_algorithm}
\end{figure}
\item One way hash function mechanism\\
Hash functions of all kinds, whether standard or lightweight, produce a fixed-lenght MD \citep{sbcw}. It is very useful for producing a hex value that cannot be decoded or used to extract the original text. The hash is evaluated based on its resistance to collisions, preimage and second preimage.Standard hash functions include MD5 and SHA-1, while lightweight hash functions include Lesamnta-LW (LLW), ARMADILLO, GLUON, LHash, and DM-PRESENT. This proposal focuses its study on the LLW algorithm because it has advantages in supporting resources, especially in resource-constrained devices. Among the methods recommended by NIST, the lightweight cryptography initiatives include LLW, PHOTON, and SPONGENT for hashing. On an 8-bit processor, LLW needs only 50 bytes of RAM (similar to AES' S-box structure), and is five times faster than SHA-256 \citep{fp75,scit}. A new family of cryptographic hash functions has been submitted to NIST as part of a competition to develop cryptographic hash algorithms. \\
LLW uses many versions of MD such as 224, 256, 384 and 512. An LLW is a plain Merkle-Damgard iterated hash algorithm, taking as input a key that is $n$/2 bits and a plaintext that is $n$ bits, where n is an even number. LLW, for instance, uses AES as its underlying cryptographic primitive, while SHA-256 uses SHACAL-2. Sensors and RFID devices that use restricted resources require security hash functions \citep{fp74,rws,rwr}. LLW uses 256-bit plaintext and 128-bit keys with an AES-based block cipher. LLW only contains the length of data input in last block, not any parts of data, in Padding. The preimage resistance of LLW is ensured by this feature. In this function, 64-round block ciphers are used. The key is 128 bits long and the input data is 256 bits long. A one-key scheduling algorithm is used in the first phase of LLW's block cipher, and a one-key mixing algorithm is used in the second phase to generate ciphertext by using data and round key. In the mix function, XOR operations, word-wise permutations, and a non-linear function $Gf. PQ = MixColumns . SubBytes$ are combined. Where $SubByte$ is a non-linear substitution operation. The input is divided into four bytes (s0, s1, s2 and s3) and then the substitution box ($Si’=S-Box (Si)$) is implemented. The $MixColumns$ operation is defined by the AES maximum distance separable matrix multiplication given by $GFe (2^8)$ \citep{fp74}. More details about this algorithm are available at \citep{fp77}. The following Tables~\ref{tab:lesamnta-performance} and \ref{tab:lesamnta-security-features} show a comparison of performance and security features between Lesamnta-LW and other hash functions\citep{fp78}.
%table
\begin{table}[t]
	\centering
	\caption{Performance comparison between Lesamnta-LW and other hashing algorithms}
		\includegraphics[width=10cm,height=7cm]{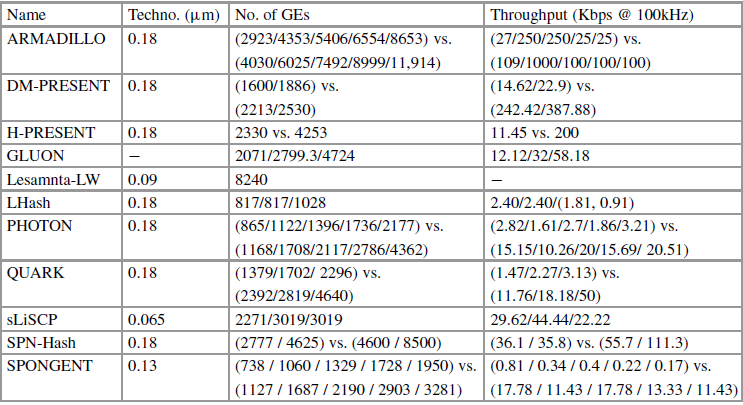}	
	\label{tab:lesamnta-performance}
\end{table}
%table
\begin{table}[t]
	\centering
	\caption{Comparison of security features between Lesamnta-LW and other hashing algorithms}
		\includegraphics[width=13cm,height=10cm]{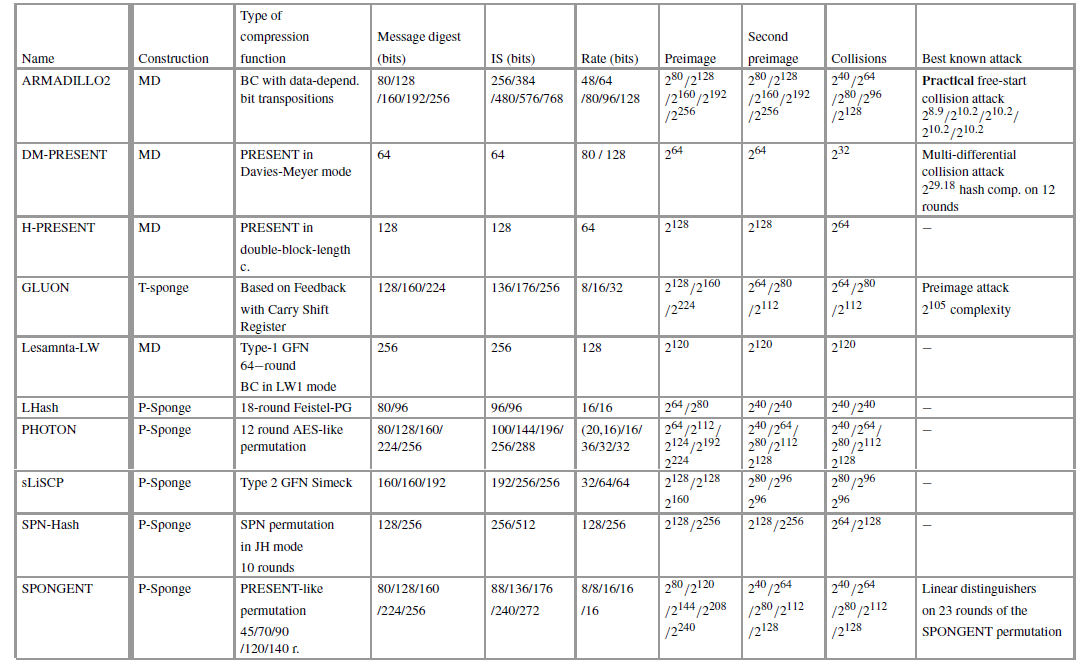}
	\label{tab:lesamnta-security-features}
\end{table}
\item Authorization control models\\
A medical records access control model is a mechanism that checks users' privileges against authorization policies to preserve the confidentiality of medical records \citep{cnac}. These privileges are specified by a database system (DBS) authorization administrator or privacy officer. These authorizations could be specified by following either a DAC policy, MAC policy, RBAC policy, or ABAC. Each authorization system requires access control model (ACMs) to authorize users' access to the medical records. Many ACMs, and each model relies on a specified manner and set of rules. Table~\ref{tab:comparison_models} shows the differences between the control models (DAC, MAC, RBAC and ABAC). However, the ACMs most used in recent research and applied in health applications are RBAC and ABAC.\\ \\

The RBAC model is used as an option substitute for the DAC and MAC models. This concept was developed by David Ferraiolo and Rick Kuhn in 1992, in which the system administrator creates roles and grants rights to those roles according to the functions performed in an organization. Data access privileges and rights are associated with each role in RBAC's system, which makes it secure as a result of its structure of assigned roles to users. This ACM categorizes users by roles such as patient, doctor, advisor, researcher ... etc., each with its own privileges and rights. Roles in the system are assigned to clients depending on their jobs. Since roles function as a connect between data access modes and clients, RBAC is better suited and less complexity for health environments than DAC and MAC \citep{fp80}.\\ \\

ABAC is one of the promising alternatives, which evaluates conditions, data attributes, and user attributes, as well as policies specifying those conditions and attributes. The ABAC model has recently drawn significant interest for protecting the privacy of medical records. This ACM utilize client attributes (like name, job, address, sex, age, marital status, phone number, location, time and health status) to authorize clients to access the server's health database more accurately (fine-grained) and privately \citep{rp}. The ABAC model proposed in 2011 to overcome the limits in the most widely famous control access models (DAC, MAC, and RBAC). Due to the wide range of attributes it addresses, ABAC is a rich model. Cloud computing, IoT, Big Data, VANETs, Internet of Things and especially healthcare all have applications where ABAC supports administration, authorization, risk intelligence, and scalability. ABAC categorizes attributes into subject, object, action, and environment. After extensive review of the ACM models, it is clear that the RBAC and ABAC models have important advantages in supporting the privacy of medical records and are well suited to this proposal \citep{fp81}.
%table
\begin{table}[t]
	\centering
	\caption{Differences between control and access models}
		\includegraphics[width=10cm,height=5cm]{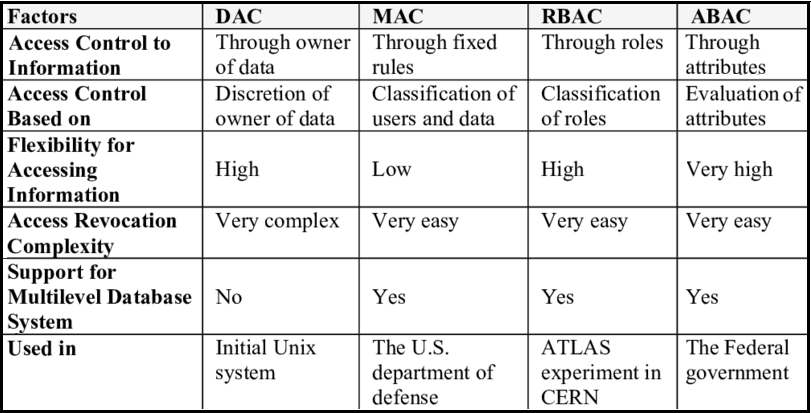}
	\label{tab:comparison_models}
\end{table}
\item Data management and control mechanism\\
In the proposed approach, the EHR database is an exceedingly important component. TAs health systems struggle to deal with a wide range of coordination for medical records, the repositories store data and information in different forms. Extensible access control (XML) is therefore suitable for online data transfer of many kinds.  XML, being a symbolic language, employs an easy and adaptable process to describe, exchange, and handle data in online settings \citep{fp82}.\\
Nevertheless, XML files should support various levels of data security for sensitive data throughout the file or in parts of it  \citep{fp83}. Access to medical records is a considerable issue in big data management systems that utilize various mechanisms. Furthermore, the  interchange of data online has become essential and requires to perform access authorization, especially in EHR approaches. Access control and XML-based data management are both covered by the extended access control markup language (XACML) standard.  At a fine-grained level, XACML offers flexible and efficient data access and authorization properties \citep{fp84}.  It contains numerous qualities that make it appropriate for online use, and it is supplied by the organization for the improvement of structured information standards (OASIS).  Policy, algorithm, attribute, numerous subjects, policy distribution, implementation independence, and duties are a few examples \citep{fp84}.\\
Prior to using units like policy enforcement points (PEP), policy decision points (PDP), policy administration points (PAP), policy information points (PIP), and policy retrieval points (PRP), XACML uses certain policies to assess access requests.  XACML operations are depicted in Figure~\ref{fig:xacml} \citep{fp82} (PEP transmits and receives requests and access responses to the database; PDP evaluates decisions; PAP creates policies based on user attributes; PIP recovers user attributes; and PRP retrieves user data from the database).  Through PEP, the subject receives the decision outcome (permit, refuse, not applicable, or indeterminate).
\begin{figure*}[t!] 
\centering
  \includegraphics[width=10cm,height=7cm]{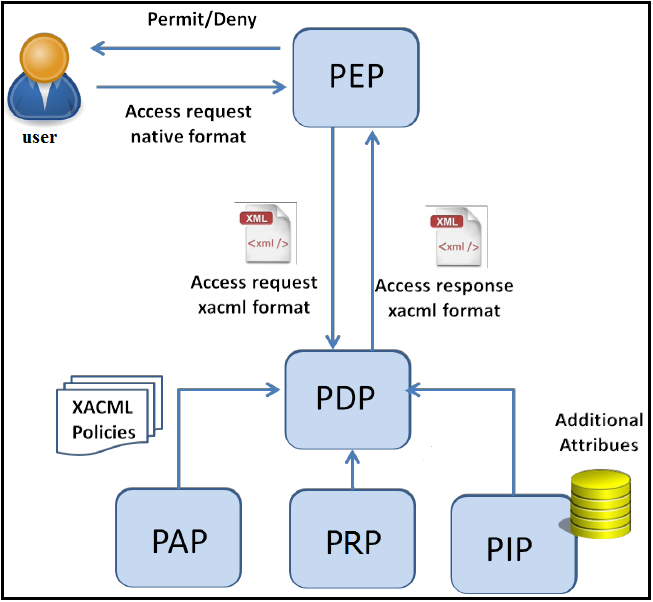}
	\caption{XACML architecture}
	\label{fig:xacml}
\end{figure*}
\item Secret sharing mechanism\\
Secret-sharing mechanisms are typical secure multi-party calculation protocols \citep{ospf}. With prime fields, it is locally leakage-resistant to arbitrary one-bit leaks from each secret share. A master secret ($MS$) is generated by a set of secret sharing ($SS_s$) and threshold ($t$) in the Shamir or secret sharing scheme ($SS_s$, $t$). Some or all of the $SS_s$ could be used to generate the master secret. This mechanism specifies the minimum number of secrets required to reconfigure $MS$. This mechanism includes the generation and reconstruction phases \citep{fp86}. Clients ($C_i$) receive one secret sharing ($SS$) from the server as part of the generation phase ($MS$) is split into a set of secrets sharing ($SS_1$, $SS_2$, etc.). Reconstruction requires $C_i$ to perform any set of secrets ($SS_s$), depending on $t$, to reconstruct $MS$ , this ensures homomorphism and correctness features. This guarantees the secrecy feature since $C_i$ cannot obtain information from the server while $t$-1 from $SS_s$.  Hackers have a hard time figuring out the $MS$, and the secrets set up for the $MS$ are anonymous \citep{fp87}; they have no way of knowing if they belong to particular individuals.  The Shamir mechanism provides an anonymous method for creating a $MS$ with a number of features, including ease of creating a $MS$ from a group of secrets, development of a new secret for one-time use, a $MS$ size equal to $C_i$s' $SS_s$ sizes, and complete security in hiding $C_i$s' $SS_s$.
Figure~\ref{fig:shamir_threshold_scheme} \citep{fp85} describes the use of the Shamir scheme to create the secret from the users' secret set. Nonetheless, Shamir mechanism is still to be vulnerable to penetration. 
\begin{itemize}
\item The wireless communication environment is considered an insecure medium for unprotected transmission of network members' secrets, which could put patient data at risk.
\item Collecting t shares to retrieve the master secret key is a complex problem in the connections of large networks with a huge number of members.
\item Assuming all network members are honest and loyal in a large network of hundreds of patients and providers is a naive strategy.
\item Distributing each member's role in helping others rebuild and recover the master secret may overburden the server due to the high communication between network members.
\item Any authorized member with access to network services and broadcast shares can reassemble the polynomial and discover the secret key. Therefore, any legitimate malicious member can supply a forged secret without anyone finding out.
\end{itemize}
\begin{figure*}[t!] 
\centering
  \includegraphics[width=10cm,height=7cm]{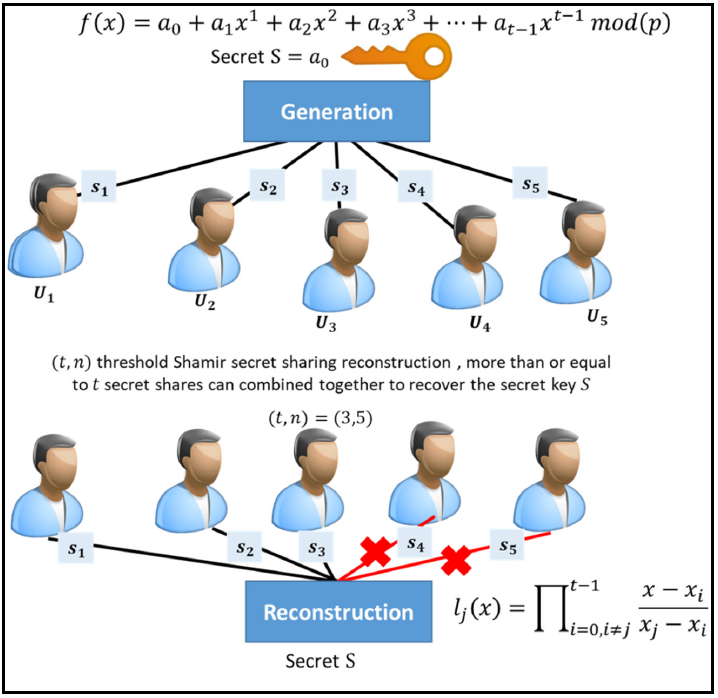}
	\caption{Shamir threshold scheme}
	\label{fig:shamir_threshold_scheme}
\end{figure*}
\end{itemize}

\section{The Recommended Approach for Authorizing Credible Sensors and Users}
\label{sec:proposed}
This section presents the technique for the suggested authorization approach that provides privacy-preserving methods to guarantee member sensors'/users' authorization in medical care apps.  The symbols used in this study are described in Table~\ref{tab3}.
\begin{table}[!t]
\begin{center} 
\caption{Manuscript's symbols} 
\label{tab3}
\scriptsize
\setlength{\tabcolsep}{2pt}
\begin{tabular}{|p{83pt}|p{171pt}|}
\hline
Notations                                                                                       & Description\\ 
\hline
    $U_{i}$, $Sen_i$; $BS$, $IS$ and  $RS$                               & User entity, Basestation, Information and Repository servers\\ 	
    $Sen_S$, $BS_S$, $IS_S$ and $RS_S$                                 & Signatures generated by $Sen$, $BS$, $IS$ and $RS$\\
    $E_{AES}$, $D_{AES}$                                                         & AES Encryption and Decryption\\	 
    $EC_{SHA}$, $EC_{LLW}$                                                   & ECDSA with SHA-1, ECDSA with Lesamnta-LW\\                                            
    $SS$                                                                                            & Shamir secret\\
    $Sen_N$, $BS_N$, $IS_N$ and $RS _N$                             & Random nonces and random secret nonce\\
    $TS_i$                                                                                        & Time stamp\\
    $V_{tm}$                                                                                   & Temporary value\\
    $Sen_{ID}$, $BS_{ID}$, $IS_{ID}$ and $RS_{ID}$         & $Sen$, $BS$, $IS$ and $RS$ identifiers\\
     $\|$, $\oplus$                                                                             & Concatenation operation, Exclusive or operation\\    
\hline
\end{tabular}
\end{center}
\end{table}
\subsection{Model of the Network}
\begin{figure*}[t!] 
\centering
  \includegraphics[width=13cm,height=6cm]{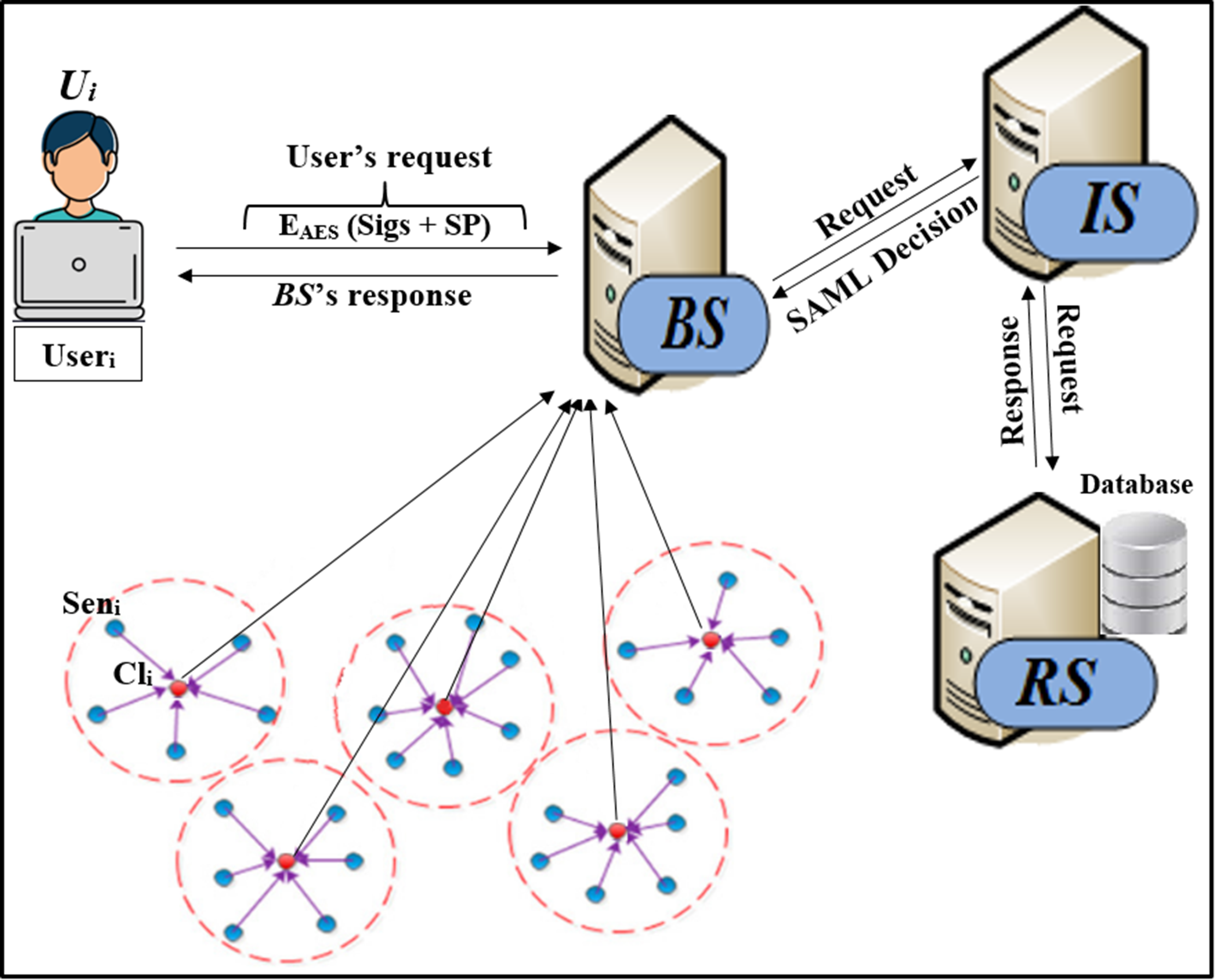}
	 \caption{Proposed authorization approach}
	\label{fig:ProposedAuthorizationApproach}
\end{figure*}

Anonymity with the SAML is an authorization method that functions with health sensor networks, as shown in Figure~\ref{fig:ProposedAuthorizationApproach}.  The user ($U_i $), sensor ($Sen_i$), cluster head ($Cl_i$), basestation server ($BS$), information server ($IS$), and repository server ($RS$) are the elements that make up the network model.  These entities interact with one another in the suggested approach to ensure authorization and safeguard the privacy of users and sensors when they access the health repository. Users and sensors are unable to connect directly to $IS$ and $RS$ due to the $BS$ gateway.  The $Sen_i$/providers information on the information server ($IS$) is kept apart from the patient dataset on $RS$.  Each $U_i$/$Sen_i$ generates an access request, which is then transmitted to the $BS$.  Once the authorization details for $U_i$ or $Sen_i$ have been verified, $BS$ forwards the authorization request to $IS$ for validation; in the event that the request is deemed invalid, $BS$ replies "reject" to $U_i$/$Sen_i$. $IS$ analyzes the permission request it receives from $BS$ after validating the access request and confirming signatures and other security parameters.  If all tests and assessments are valid, $IS$ requests that $RS$ retrieve/store patient data; if not, $IS$ replies to $BS$ with a "reject" message.  This means that $RS$ looks for security parameters (SP) and signatures (Sigs). If everything has been done correctly, $RS$ then transmits the required information to $IS$, which in turn sends the "accept" answer to $U_i$/$Sen_i$ by $BS$ to permit access to and storage of the medical records. The approved $U_i$/$Sen_i$ will receive the "accept" answer and a copy of the required data.  To provide a high degree of $U_i$/$Sen_i$ privacy, this method concentrates on protecting/storing data and requests.  The open-source project that uses SAML v2.0 is responsible for protecting patient data confidentiality.
% Define shape styles for tikz package with flowchart
\tikzstyle{decision} = [diamond, draw, fill=white!20, text width=4.5em, text badly centered, node distance=3cm, inner sep=0pt]
\tikzstyle{block} = [rectangle, draw, fill=white!20, text width=9em, text centered, minimum height=3em ,minimum width=4em]
\tikzstyle{line} = [draw, -latex']   %[draw, -triangle 45]
\tikzstyle{cloud} = [draw, ellipse,fill=white!20, node distance=3cm, minimum height=2em]
\subsection{Developing the Suggested Authorization Strategy}
In this section, we shall outline the privacy implications of the suggested permission strategy.
\begin{itemize}
\item \textbf{Combining the signatures of ECDSA, and Lesamnta-LW}\\
To ensure that security standards are satisfied with respect to the integrity of patient data, the suggested authorization mechanism uses ECDSA-256 in conjunction with policies and requests.  We have implemented ECDSA signatures with senders’ and receivers’ information to ensure the integrity property, which prevents changing information in policies and requests; the authentication property, which prevents external hackers; and the non-repudiation property, which prevents authorized sensors/users from denying their requests to receive medical data. Implementing security criteria is essential in systems that handle sensitive data, such health sensor systems.  While the $U_i$/$Sen_i$ sign the request using parameters ($Sen_N$ and $BS_N$), the servers ($BS$ and $IS$) check the request's Sigs.  The $IS$ forwards the request to the SAML v2.0 evaluation procedure if it is valid; if not, it is rejected.  The suggested authorization method uses ECDSA's Sigs to partially hide the $Sen_i$ and $BS_i$ information when sending and receiving SAML requests. This algorithm's appropriate performance and security level make it acceptable for usage in large systems that leverage health sensor technologies.  We employed Lesamnta-LW in ECDSA rather than SHA1 to assist the security and performance of our approach, which has a good impact on health sensors.  Lesamnta-LW is utilized in ECDSA because it supports ECDSA signatures to increase security and offers superior performance when working with sensors.
\item \textbf{Administration of policies in the suggested Approach}\\
By using SAML, the system administrator is in charge of constructing policies for patients and healthcare providers in $IS$. SAML policies define the rules and conditions for how SAML assertions should be handled and enforced within an authentication and authorization system. SAML policies:
\begin{itemize}
\item Authentication policies: These policies specify the requirements for authenticating sensors/users before granting access to a service. They can include factors such as username/password authentication, multi-factor authentication (MFA), or integration with external identity providers for federated authentication.
\item Attribute release policies: These policies determine which attributes of a user's/sensor's identity should be included in the SAML assertion sent from the IdP to the SP. Attribute release policies often involve mapping and transformation of user attributes between different identity systems.
\item Authorization policies: These policies define the rules for granting or denying access to specific resources or services based on the information contained in the SAML assertion. They can be based on attributes such as user/sensor roles in the NGAC model, group memberships, or custom attributes defined in the SAML assertion.
\end{itemize}
Figure~\ref{fig:policy} displays the types of policies and authorization methods. The policy ID, sender, recipient, and implementation rules for the policy make up the proposed authorization strategy. The creation of datasets for information for all sensors and users is the initial step in the proposed authorization approach. Policies are established based on prior datasets. Policies on the server are shielded against malicious attacks by using signature-based policy protection. This policy may include a variety of specifications, such as how to choose the day and time of data access, the time limit for a certain day, or the total number of accesses.
\begin{figure*}[t!] 
\centering
  \includegraphics[width=11cm,height=4cm]{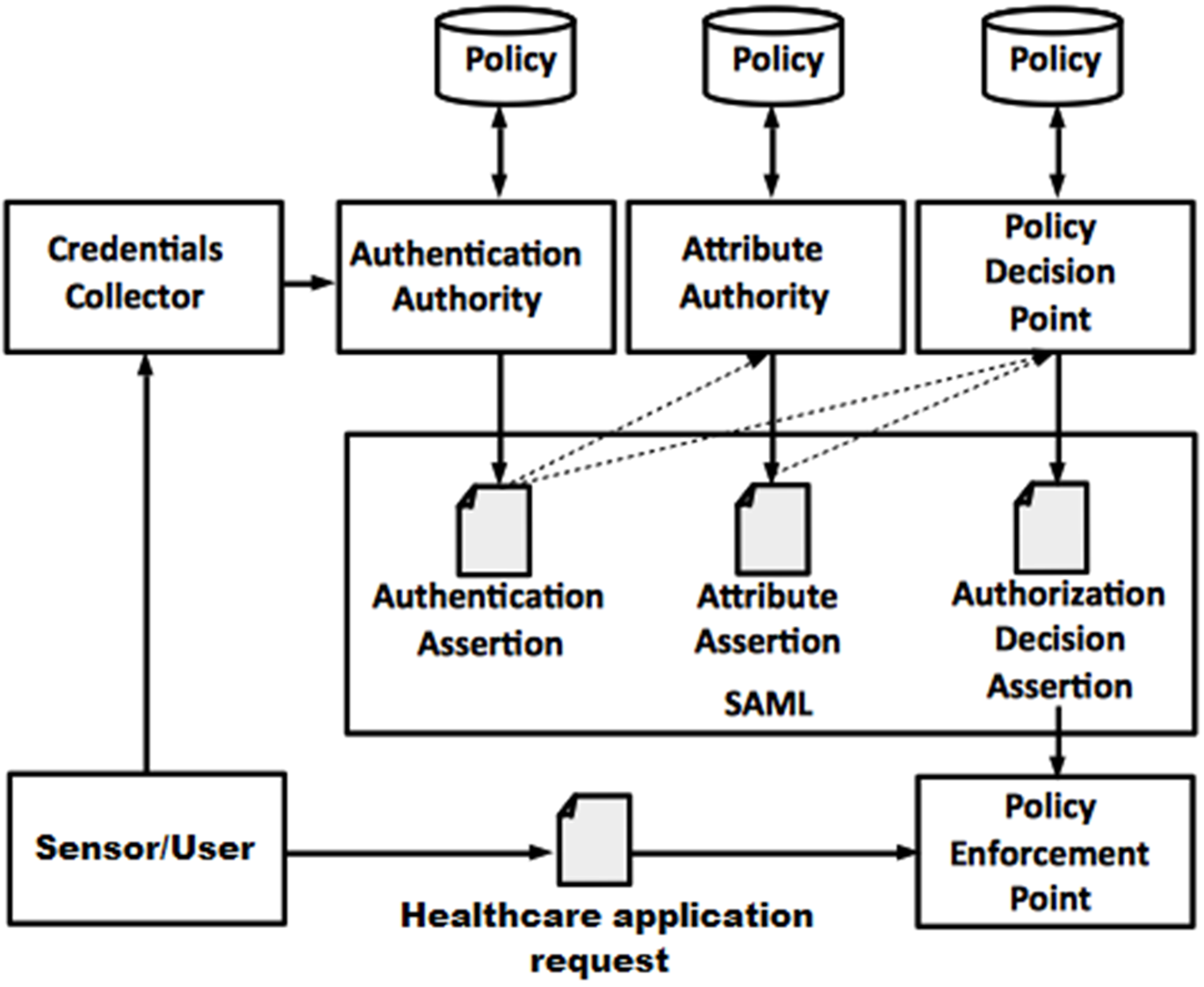}
	\caption{Proposed authorization approach policy}
	\label{fig:policy}
\end{figure*}
\item \textbf{Requests and responses between sensors/users  and servers}\\
The proposed authorization approach requires sensors and users to create an authorization request before they may access medical records. The sender and receiver details are included in this request. The parameters of the $SS||Sen_N||Sen_ID$ are used as a piece of single information by the $Sen_i$ application to create the ECDSA's Signature for the senders and receivers. Additionally, after confirming the Signatures, the $Sen_i$ application utilizes a portion of the $Sen_ID$ to explain to the $BS$ the identification of the entity in order to decide the desired policy. The request is then forwarded by the $Sen_i$ via the $BS$ to the $IS$ for analysis. The request is evaluated by the $IS$, and the $BS$ then provides the result (permit or refuse) to the $Sen_i$. In practice, SAML request files are typically generated and managed by the SAML implementation in our proposed approach. They are encoded and transmitted to the destination using the specified binding method, such as a form submission. The process of the request authenticates the sensor/user and generates a SAML response containing the requested assertions and attributes.
\item \textbf{Utilizing Shamir mechanism}\\
We adopted the Shamir method to the proposed authorization strategy to boost the security of sensors and users. Due to the privileges assigned to legitimate devices, $Sen_i$s are real devices that are capable of posing a risk inside. The proposed authorization approach signs all healthcare sensors' and users' signatures using ECDSA to produce a Shamir secret. Then, the proposed authorization approach creates secrets sharing ($SS_s$) from an $SS$ using the Shamir manner. The $SS_i$ is delivered to each network device through a secure communication channel. To rebuild SS, $Sen_i$ needs a collection of $SS_s$. Threshold = 3 is used in the proposed authorization strategy, which indicates that the randomly chosen $SS_s$ need at least 3 $SS_s$ to create $SS$.  Additionally, based on $V_ tm$, $IS$ identifies the network device as being valid, verifies the original $SS$ using the Shamir secret and ECDSA's signature, and then assesses the request using security parameters. By combining Shamir's method with SAML, the authenticity feature is added since a $Sen_i$ cannot access data using the same $SS_s$. This procedure makes it possible for the proposed authorization approach to ensure patient data privacy and safeguard patient data from serious threats.
\item \textbf{Applying encryption on authorization request}\\
In our protocol, we rely on AES-256 with 10 rounds to mask authorization requests to store health sensor data to protect it from attacks. The use of a symmetric encryption algorithm will enable the health sensors to operate with high performance as well as support the security of network parameters and data collected by the health sensors. Our approach uses AES-256 to protect information signed by ECDSA-Lesamnta-LW and SAML authorization requests.
\subsection{Proposed authorization approach protocol}
This section will go through the proposed authorization protocol framework for authorizing sensors and users in detail. The protocol request contains $SP$ for a subject (the sender) and an object (the receiver). Figure~\ref{fig:Proposed_authorization_protocol} displays the authorization processing.
\begin{itemize}
\item Initially, the $Sen_i$ device retrieves the $Sen_{ID}$ and $BS_{ID}$ identifiers for use in access operations and stores the data collected by the sensors in $RS$. Then $Sen_i$ generates a random number $Sen_N$, timestamp ($TS_i$) and uses Shamir secret to produce the original $SS$. Next, $Sen_i/U_i$ performs an electronic signature ($Sen_S$) with $EC_{LLW}$ to ensure the integrity of the privacy request and based on the $SS||Sen_N||Sen_{ID}$ security parameters. Then $Sen_i$ performs the masking of $TS_i$ and $BS_{ID}$ in a temporary value of $V_{tm}$ to prevent information from being leaked. Then $Sen_i$ uses SAML to generate a privacy request that includes hidden privacy information for the sending and receiving device. Finally, $Sen_i$ uses AES-192 to protect the SAML request, sign $Sen_S$, random number $Sen_N$ and then transferred the encrypted request $R_{Sen_1}$ to $BS$.
\item $BS$ gets the AES-192 encrypted SAML request and performs decryption of the $R_{BS_1}$ request. Then $BS_{ID}$ and $R_{BS_1}$ are used to retrieve the random number $Sen_N$ which is used with $Sen_{ID}$ and $SS$ in signature processing to find $BS_{S_1}$ and tested with $Sen_S$ received. After then, $BS$ verifies $TS_i$ to check if the request was delivered on time. Next, $BS$ prepares the SAML authorization request containing $V_{tm}$ and an electronic signature $BS_{S_2}$ to be encrypted and sent to the $IS$ information server.
\item $IS$ receives a SAML authorization request containing a $V_{tm}$ timer value, $BS_{S_2}$ signature, $TS_i$ send time, as well as some other $SP$ security parameters such as $IS_{ID}$ and $SS$. $IS$ verifies the integrity of the information with $IS_{S_1}$=$BS_{S_2}$ and does not delay the authorization request by $TS_i$. To authorize the request in $RS$, $IS$ computes a secret value ($Sec$) which includes $Sen_{ID}$, $BS_{ID}$, $IS_{ID}$ and $RS_{ID}$ which is used to authorize requests in the next connections. Similarly, $IS$ computes the SAML authorization request containing $V_{tm}$, $IS_{S_2}$ and then $R_{IS_2}$ encrypts it and sends it to $RS$ to prove access and store the data collected by $Sen_i$.
\item At this juncture, $RS$ gets the SAML authorization message from $IS$. $RS$ validates the SAML request, $RS_{S_1}$ and $TS_i$. If all parameters are valid, $RS$ allows access or storage of the data collected by $Sen_i$, otherwise, it refuses the connection.
\item Finally, $RS$ generates a response request that includes a token accept request ($T_{AR}$) that contains a signature, a random number and the secret value $Sec$. Then $RS$ encrypts the request ($R_{RS_2}$) by $E_{AES}(T_{AR}||TS_i||RS_{ID})$ and sends it to $IS$ and then to $BS$ until it reaches $Sen_i $ which decrypts the request and verifies the signature and then stores $Sec$ for next incoming connections.
\end{itemize}
\begin{figure*}[t]
\centering
\scriptsize 
\begin{tikzpicture}
\scalebox{0.95}{ %To reduce font
\notsotiny

%To draw vertical dashed line
\draw [dashed] (4.5,1) -- (4.5,10.8);
\draw [dashed] (8.3,1) -- (8.3,10.8);
\draw [dashed] (12.0,1) -- (12.0,10.8);

%To write entities 
\node [right=0.3cm] at (25pt,325pt){\Large\textbf{\fbox{$Sen_i/U_i$}}};
\node [right=4.8cm] at (25pt,325pt){\Large\textbf{\fbox{$BS$}}};
\node [right=9.0cm] at (25pt,325pt){\Large\textbf{\fbox{$IS$}}};
\node [right=12.5cm] at (25pt,325pt){\Large\textbf{\fbox{$RS$}}};

% Seni/Useri
\node [right=-0.1cm] at (25pt,300pt){\textbf{Retrieves $Sen_{ID}$ and $BS_{ID}$}};
\node [right=-0.1cm] at (25pt,290pt) {Generates  $Sen_{N}$ and $TS_{i}$};
\node [right=-0.1cm] at (25pt,280pt) {Generates $SS$};
\node [right=-0.1cm] at (25pt,270pt) {$Sen_S=EC_{LLW}(SS||Sen_N||Sen_{ID})$};
\node [right=-0.1cm] at (25pt,260pt) {$V_{tm}=Sen_S\oplus BS_{ID} \oplus TS_i$};
\node [right=-0.1cm] at (25pt,250pt) {Creates SAML ($V_{tm}$) request};
\node [right=-0.1cm] at (25pt,240pt) {$R_{Sen_1}=E_{AES}(SAML (V_{tm}||Sen_{S}||Sen_N)$};
\node [right=0.6cm] at (25pt,230pt) {$\oplus BS_{ID}$};

% To draw arrow with text
\draw [->,>=stealth] (1.2,7.2) -- (4.2,7.2) node[above,pos=0.37] {\textbf{Sends $R_{Sen_1}$ request}};

\node [right=-0.1cm] at (25pt,70pt) {$R_{Sen_2}=D_{AES}()$};
\node [right=-0.1cm] at (25pt,60pt) {Checks $TS_i$};
\node [right=-0.1cm] at (25pt,50pt) {Extracts $Sec$};
\node [right=-0.1cm] at (25pt,40pt) {Checks $BS_{S_1}=Sen_S$};
\node [right=-0.1cm] at (25pt,30pt) {Stores $Sec$ for next connections};

%BaseStation Server side:
\node [right=3.6cm] at (25pt,290pt) {\textbf{Receives SAML's request}};
\node [right=3.6cm] at (25pt,280pt) {Retrieves $BS_{ID}$};
\node [right=3.6cm] at (25pt,270pt) {$R_{BS_1}=D_{AES}(SAML(V_{tm})||Sen_S||Sen_N)$};
\node [right=4.3cm] at (25pt,260pt) {$\oplus BS_{ID}$};
\node [right=3.6cm] at (25pt,250pt) {$Sen_N=R_{BS_1}\oplus BS_{ID}$};
\node [right=3.6cm] at (25pt,240pt) {$BS_{S_1}=EC_{LLW}(SS||Sen_N||Sen_{ID})$};
\node [right=3.6cm] at (25pt,230pt) {Checks $BS_{S_1}=Sen_S$};
\node [right=3.6cm] at (25pt,220pt) {Extracts and checks $TS_i$};
\node [right=3.6cm] at (25pt,210pt) {Retrieves $IS_{ID}$};
\node [right=3.6cm] at (25pt,200pt) {$BS_{S_2}=EC_{LLW}(SS||IS_{ID})$};
\node [right=3.6cm] at (25pt,190pt) {Generates $TS_i$};
\node [right=3.6cm] at (25pt,180pt) {$V_{tm}=BS_{S_2}\oplus IS_{ID}\oplus TS_i$};
\node [right=3.6cm] at (25pt,170pt) {Create SAML ($V_{tm}$) request};
\node [right=3.6cm] at (25pt,160pt) {$R_{BS_{2}}=E_{AES}(SAML(V_{tm})||BS_{S_2})\oplus IS_{ID}$};

%To draw arrow with text
\draw [->,>=stealth] (5,5.0) -- (8,5.0) node[above,pos=0.37] {\textbf{Sends $R_{BS_{2}}$ request}};

\node [right=3.6cm] at (25pt,110pt){\textbf{Creates SAML's response}};
\node [right=3.6cm] at (25pt,100pt) {$R_{BS_{3}}=D_{AES}()$};
\node [right=3.6cm] at (25pt,90pt) {Checks $TS_i$};
\node [right=3.6cm] at (25pt,80pt) {Extracts $Sec$};
\node [right=3.6cm] at (25pt,70pt) {Generates $BS_{N}$ and $TS_i$};
\node [right=3.6cm] at (25pt,60pt) {Computes $T_{AR}=BS_{S_1}\oplus BS_N\oplus Sec$};
\node [right=3.6cm] at (25pt,50pt) {$R_{BS_{3}}=E_{AES}(T_{AR}||TS_i||BS_{ID})$};

%To draw arrow with text
\draw [->,>=stealth] (8.2,1.0) -- (5.2,1.0) node[above,pos=0.37] {\textbf{Sends $R_{BS_{3}}$ response}};

%Information Server side:
\node [right=7.4cm] at (25pt,280pt){\textbf{Receives SAML's request}};
\node [right=7.4cm] at (25pt,270pt) {Retrieves $IS_{ID}$};
\node [right=7.4cm] at (25pt,260pt) {$R_{IS_1}=D_{AES}(SAML(V_{tm})||BS_{S_2})\oplus IS_{ID}$};
\node [right=7.4cm] at (25pt,250pt) {$IS_{S_1}=EC_{LLW}(SS||IS_{ID})$};
\node [right=7.4cm] at (25pt,240pt) {Checks $IS_{S_1}=BS_{S_2}$};
\node [right=7.4cm] at (25pt,230pt) {Checks $TS_i$};
\node [right=7.4cm] at (25pt,220pt) {Retrieves $BS_{ID}$ and $RS_{ID}$};
\node [right=7.4cm] at (25pt,210pt) {$Sec=Sen_{ID}\oplus BS_{ID}\oplus IS_{ID}\oplus RS_{ID}$};
\node [right=7.4cm] at (25pt,200pt) {$IS_{S_2}=EC_{LLW}(SS||RS_{ID})$};
\node [right=7.4cm] at (25pt,190pt) {Generates $TS_i$};
\node [right=7.4cm] at (25pt,180pt) {$V_{tm}=IS_{S_2}\oplus IS_{ID}\oplus TS_i$};
\node [right=7.4cm] at (25pt,170pt) {Creates SAML ($V_{tm}$) request};
\node [right=7.4cm] at (25pt,160pt) {$R_{IS_{2}}=E_{AES}(SAML(V_{tm})||IS_{S_2}||Sec)$};
\node [right=8.1cm] at (25pt,150pt) {$\oplus RS_{ID}$};

%To draw arrow with text
\draw [->,>=stealth] (9.0,4.7) -- (11.5,4.7) node[above,pos=0.37] {\textbf{Sends $R_{IS_{2}}$ request}};

\node [right=7.4cm] at (25pt,120pt){\textbf{Creates SAML's response}};
\node [right=7.4cm] at (25pt,110pt) {$R_{IS_{3}}=D_{AES}()$};
\node [right=7.4cm] at (25pt,100pt) {Checks $TS_i$};
\node [right=7.4cm] at (25pt,90pt) {Extracts $Sec$};
\node [right=7.4cm] at (25pt,80pt) {Generates $IS_{N}$ and $TS_i$};
\node [right=7.4cm] at (25pt,70pt) {Computes $T_{AR}=IS_{S_1}\oplus IS_N\oplus Sec$};
\node [right=7.4cm] at (25pt,60pt) {$R_{IS_{3}}=E_{AES}(T_{AR}||TS_i||IS_{ID})$};

%To draw arrow with text
\draw [->,>=stealth] (11.5,1.3) -- (9.0,1.3) node[above,pos=0.37] {\textbf{Sends $R_{IS_{3}}$ response}};

%Repository Server side:
\node [right=11.1cm] at (25pt,270pt){\textbf{Receives SAML's request}};
\node [right=11.1cm] at (25pt,260pt) {Retrieves $RS_{ID}$};
\node [right=11.1cm] at (25pt,250pt) {$R_{IS_1}=D_{AES}(SAML(V_{tm})||IS_{S_2}||Sec)\oplus RS_{ID}$};
\node [right=11.1cm] at (25pt,240pt) {Extracts $Sec$};
\node [right=11.1cm] at (25pt,230pt) {$RS_{S_1}=EC_{LLW}(SS||RS_{ID})$};
\node [right=11.1cm] at (25pt,220pt) {Checks $RS_{S_1}$};
\node [right=11.1cm] at (25pt,210pt) {Checks $TS_i$};
\node [right=11.1cm] at (25pt,200pt) {If $RS_{S_1}$ and $TS_i$ is validated};
\node [right=11.1cm] at (25pt,190pt) {Then $RS$ accepts access and store};
\node [right=11.1cm] at (25pt,180pt) {Else $RS$ rejects connection};

\node [right=11.1cm] at (25pt,120pt){\textbf{Creates SAML's response}};
\node [right=11.1cm] at (25pt,110pt) {Generates $RS_{N}$ and $TS_i$};
\node [right=11.1cm] at (25pt,100pt) {Computes $T_{AR}=RS_{S_1}\oplus RS_N\oplus Sec$};
\node [right=11.1cm] at (25pt,90pt) {$R_{RS_{2}}=E_{AES}(T_{AR}||TS_i||RS_{ID})$};

%To draw arrow with text
\draw [->,>=stealth] (15.5,2.5) -- (12.5,2.5) node[above,pos=0.37] {\textbf{Sends $R_{RS_{2}}$ response}};
}
\end{tikzpicture}
	\caption{Proposed authorization protocol}
	\label{fig:Proposed_authorization_protocol}
\end{figure*}
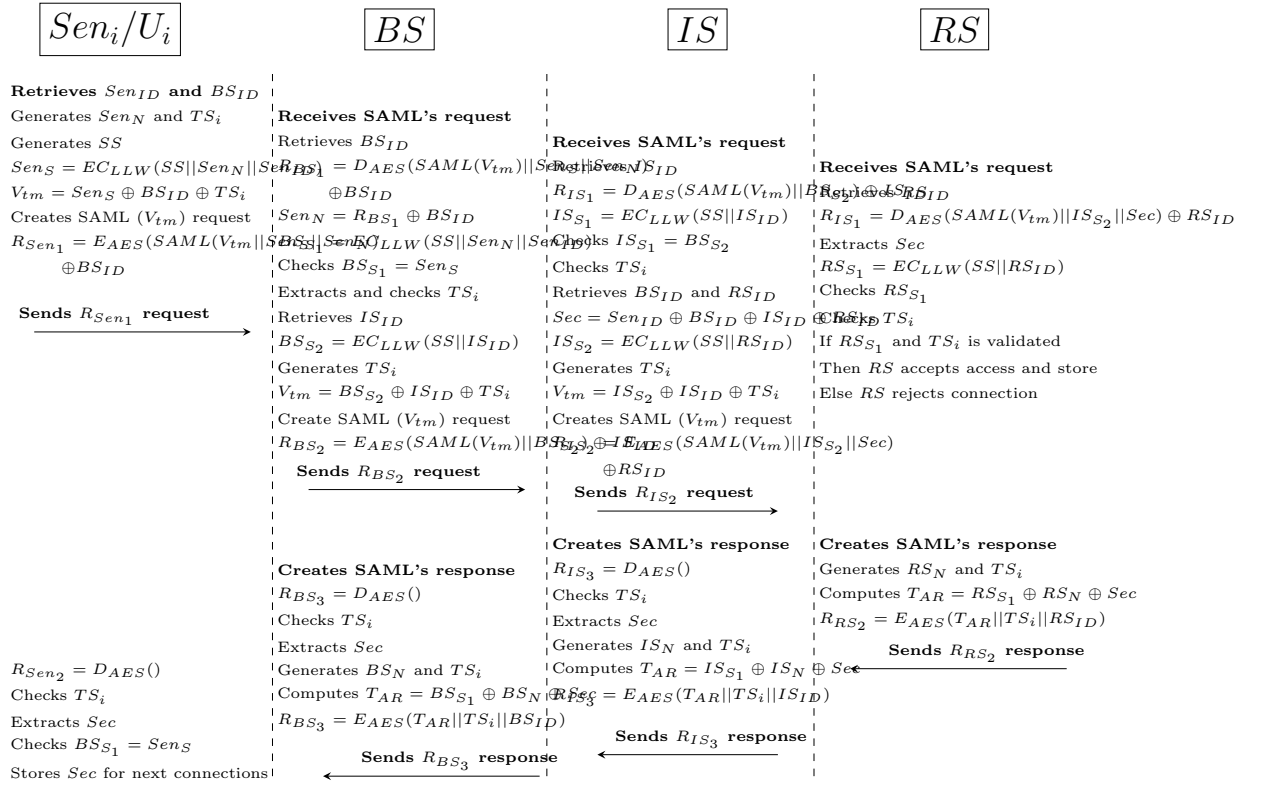
\end{itemize}
\section{Discussions and Security/Performance Analysis}
\label{sec:security_analysis}
This section investigates an evaluation of theoretical attacks and a security comparison.
\subsection{Security analysis}
By presenting hypotheses and proofs, this section will demonstrate how the proposed approach offers a high level of security against various assaults. Table~\ref{tab:security_comparison} provides a comparison of privacy properties.
\begin{itemize}
\item \textbf{Resist against a node outage attack.}\\
\textbf{Proof 1:} This threat entirely disables the operation of any wireless sensor components, including sensor nodes, communication links, and master nodes. As a result, the connection to other cluster head nodes located in other areas is severed. The attacker is trying to disconnect a specific $Sen_i$ or $Cl_i$. In our protocol, the data collected by sensors does not depend only on a particular $Sen_i$ or $Cl_i$. Any authorization request that does not contain a valid $Sen_{ID}$ and $Sen_{N}$ is rejected by $BS$ and $IS$ servers. This shows that our protocol is resistant to attacks. 
\item \textbf{Resist against the MitM attack.}\\
\textbf{Proof 2:} Consider a scenario in which a hacker tries to intercept encrypted authorization requests from network entities (such $R_Sen_1$, $R_BS_2$, and $R_IS_2$) and then replaces or alters this information of requests with his or her own messages to forward to the health sensor network. Nonetheless, the hacker is unable to amend the requests that were sent to $Sen_i$, $BS$, and $IS$ because, first, $EC_LLW$ signatures prohibit $SP$ from being altered. The alteration of requests between $Sen_i$, $BS$, $IS$, and $RS$ is also prevented by mutual authentication that includes ECDSA-Lesamnta-LW. Consequently, the proposed approach successfully repels the MitM attack.
\item \textbf{Resist against impersonation attack.}\\
\textbf{Proof 3:} Suppose a hacker impersonates a valid sensor's/user's authorization request in an attempt to get access to the network. Due to the total concealment of the security settings, this hacker is not able to construct $TS_i$ and $Sen_N$. By altering the signatures, the hacker also tries to pass as the $Sen_i$/$U_i$ device to access the network. The $Sen_S$ varies in each connection based on the $Sen_N$, rendering this scenario infeasible. As an outcome, our protocol deters threats made in the form of a false identity.
\item \textbf{Resist against rushing attack.}\\
\textbf{Proof 4:} The hacker breaks down $Sen_i$/$U_i$ from the WSN on-demand routings in an effort to make system resources scarce. He/she then swiftly broadcasts bogus route request advertisements across the WSN. As a result, this kind of threat has a major impact on a network connection and degrades networking capabilities including message delivery and control. The malicious device rushes to transfer the messages from the neighbor to the other destination sensor through a separate tunnel. Our protocol provides a countermeasure of this threat is that entities such as ($Sen_i$, $Cl_i$, $BS$, $IS$ and $RS$) do not allow the connection to be accepted until all security parameters in $V_{tm}$ are met, the hacker cannot change the routing path. Therefore, our approach resists this onslaught.
\item \textbf{Resist against vampire attack.}\\
\textbf{Proof 5:}  This threat is the type of denial of service (DoS) threat that consumes the energy of $Sen_i$ and completely declines the network. It employs the sending of a request that utilizes more network energy than a legitimate $Sen_i$ would if it sent data of a similar size to the same destination. Also, Vampires use protocol-compliant data. To counteract the vampire, our protocol introduces authentication and verification (such as  $SS$ and $TS_i$) before accepting authorization requests and prevents heavy repetitive (infinitely looped) transfers from a given entity.
\item \textbf{Resist against neglect and greed attack.}\\
\textbf{Proof 6:} By delivering the request to the incorrect sensor, the hacker chooses the longest route to convey the information. The unauthorized $Sen_i$ neglects to transfer the information and drops them randomly. Instead, hacker becomes greedy and transfer their own information to the other sensors.  This behavior will reduce the remaining energy of the sensors found in that path, thus breakdown the health sensor network quickly. Our approach uses authorization mechanisms (such as SAML authorization) to limit and detect this type of attack both between sensor nodes and all network entities.
\item \textbf{Resist against packet drop attack.}\\
\textbf{Proof 7:}  Launching a packet-dropping threat is for a malicious sensor to get involved during path formation. This is exploiting the vulnerabilities of the routing protocols utilized in health sensors which are designed based on the assumption of trustworthiness between sensors in a network. This type of attack can cause packets not to arrive in a timely manner or not to receive a response to authorization requests from entities. Our approach uses $TS_i$ to make sure authorization requests arrive in a timely manner. Also, any authorization request that does not contain $Sec$ or that does not reach the target destination indicates that the authorization request has been subjected to a drop packet attack. Therefore, our protocol detects and mitigates the risk of this attack.
\end{itemize}
 \begin{table*}[!t]
\begin{center} 
\caption{Comparison of privacy properties} 
\label{tab:security_comparison} % \notsotinynew 
\scriptsize
\setlength{\tabcolsep}{2pt}
\begin{tabular}{|p{63pt}|p{40pt}|p{30pt}|p{30pt}|p{36pt}|p{30pt}|p{34pt}|p{29pt}|p{22pt}|p{42pt}|}
\hline
Privacy property              &\citet{cp88}      &\citet{fp68} &\citet{fp69} &\citet{fp64} &\citet{fp61}  &\citet{fp55}   &\citet{cp89}    &\citet{cp90}     & \textbf{Proposed authorization protocol} \\
\hline
Anti node outage            &\checkmark       &                      &                      &                      &                       &                       &\checkmark     &                           &\checkmark\\
Anti MitM                       &                          &\checkmark  &\checkmark   &\checkmark  &\checkmark  &                       &                         &                          &\checkmark\\
Anti impersonation          &                          &                      &\checkmark   &                      &\checkmark   &                     &\checkmark      &                           &\checkmark\\
Anti rushing                      &\checkmark      &                      &                      &\checkmark  &                       &                       &\checkmark     &                           &\checkmark\\
Anti vampire                      &\checkmark      &                      &                      &\checkmark  &                       &\checkmark   &\checkmark     &                           &\checkmark\\
Anti neglect and greed    &\checkmark      &                      &                      &                      &                       &                        &\checkmark    &\checkmark       &\checkmark\\
Anti  packet drop             &                           &                      &                      &                      &                        &                      &                         &\checkmark       &\checkmark\\
\hline
\end{tabular}
\end{center}
\end{table*}
\subsection{Performance analysis}
\label{sec:performance_theoretical_analysis}
\begin{figure*}[t!] 
\centering
  \includegraphics[scale=0.39,width=10cm,height=3cm]{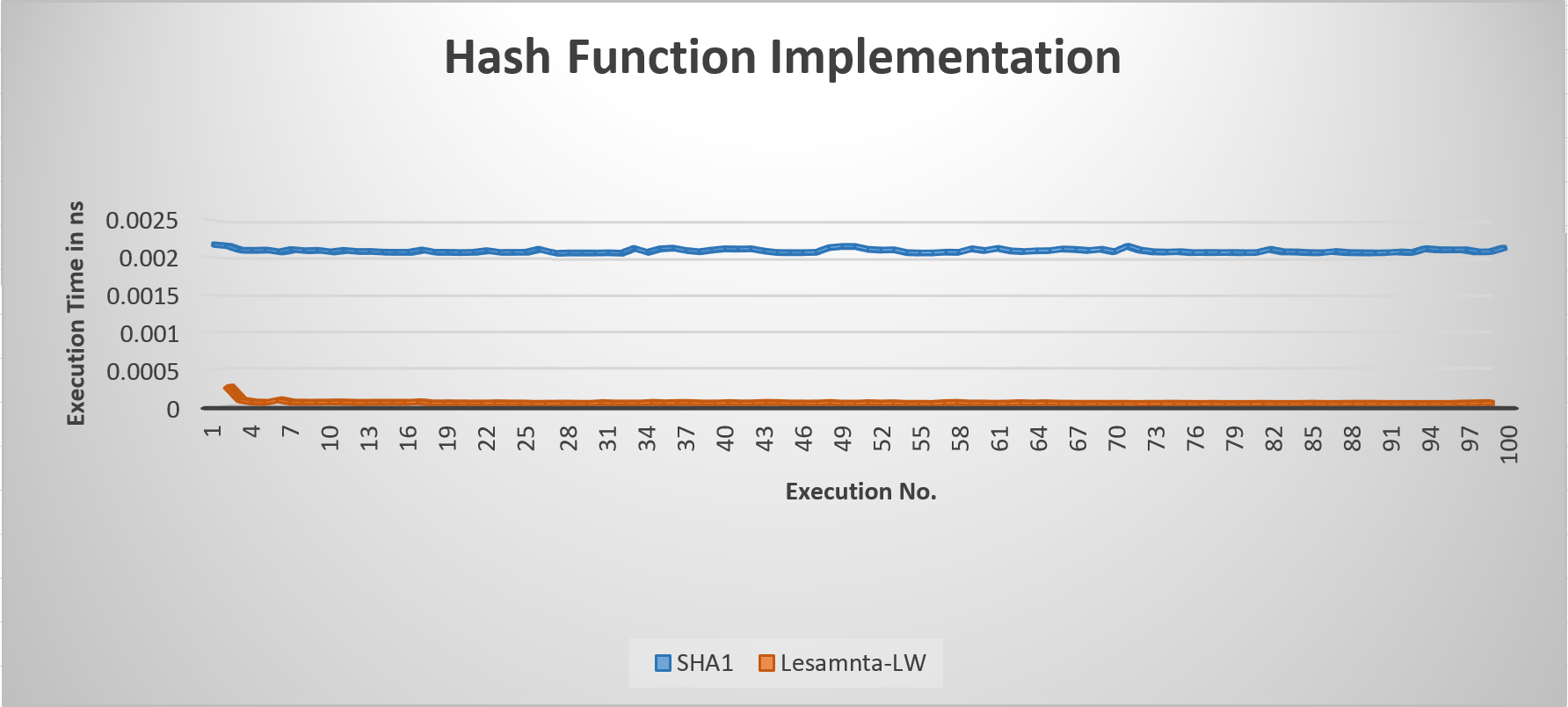}
	\caption{Comparison between SHA1 and Lesamnta-LW}
	\label{fig:sha1_lesamnta-lw}
\end{figure*}

\begin{figure*}[t!] 
\centering
  \includegraphics[scale=0.39,width=10cm,height=3cm]{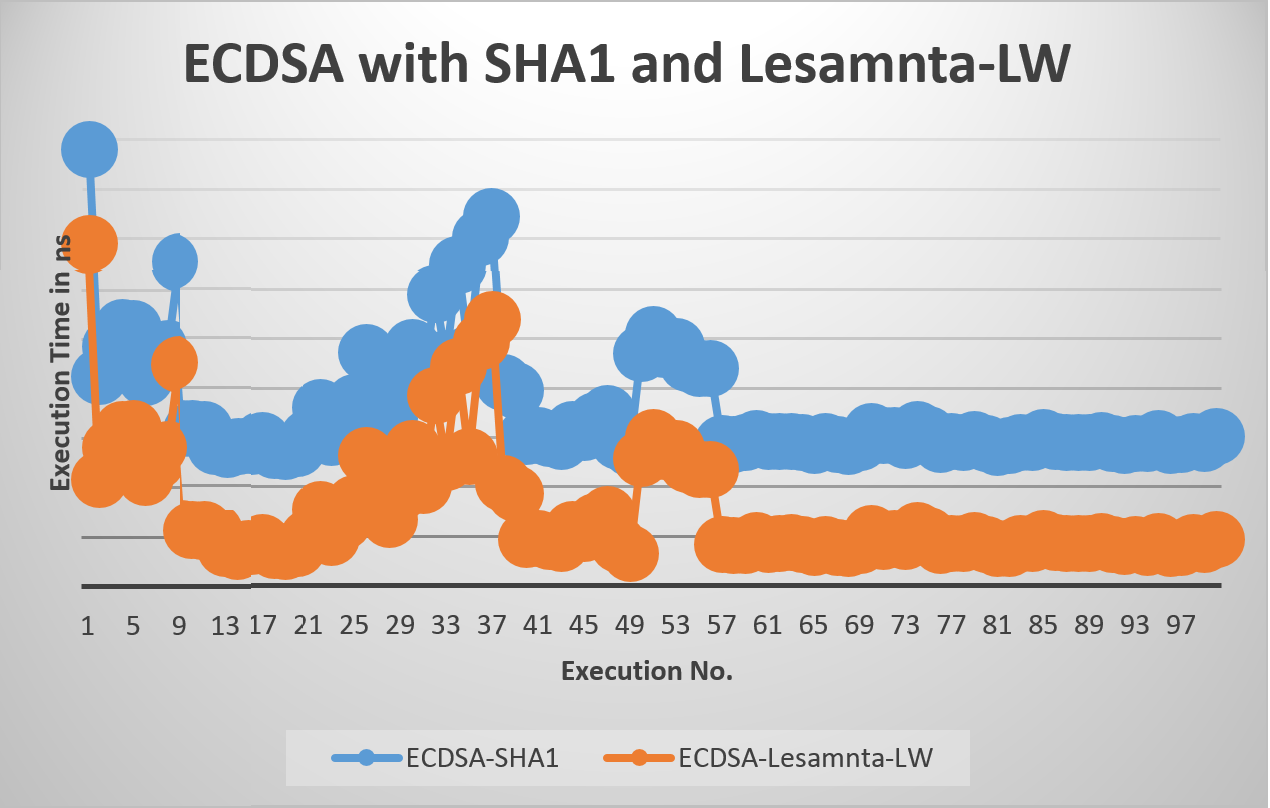}
	\caption{Comparison between ECDSA-SHA1 and ECDSA-Lesamnta-LW}
	\label{fig:ecdsa-sha1-lesamnta-lw}
\end{figure*}

\begin{figure*}[t!] 
\centering
  \includegraphics[scale=0.39,width=11cm,height=3cm]{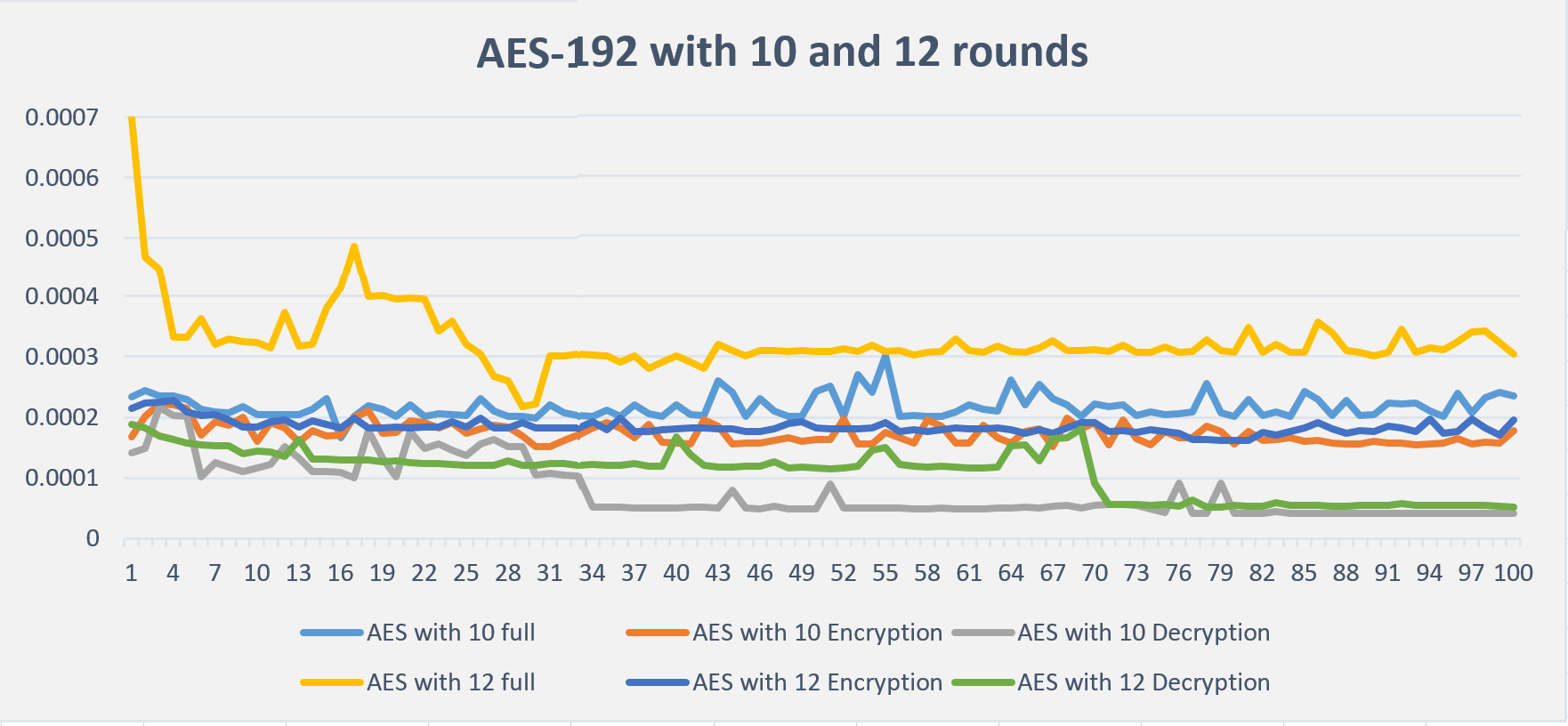}
	\caption{Performance comparison between AES with 10 rounds and AES with 12 rounds}
	\label{fig:aes_10_12}
\end{figure*}

This section describes how the proposed protocol performs in a scenario with health sensors. It offers evaluations for the 256-bit Lesamnta-LW hash function, 256-bit ECDSA signature technique, and 192-bit AES encryption cryptography. To determine the effectiveness of the proposed strategy, communication costs (storage overheads) and computation (execution time) are also considered. C programming language is used to create the application codes $Sen_i$, $BS$, $IS$, and $RS$. The findings are also applied with Ubuntu 20.04 LTS, an Intel Core i5 CPU running at 2.6GHz, a 64-bit operating system, 16 GiB of memory, and a 32.0 GB hard drive. Our protocol performs best as can be seen in Figure~\ref{fig:sha1_lesamnta-lw} which shows Lesamnta-LW outperforming SHA1. Also, Figure~\ref{fig:ecdsa-sha1-lesamnta-lw} shows the superior performance of ECDSA-Lesamnta-LW over ECDSA-SHA1. Finally, Figure~\ref{fig:aes_10_12} shows that AES 192 bits with 10 rounds performs better than AES with 12 rounds.
 Table ~\ref{tab:performance_comparison} compares performance of computation cost our protocol with existing protocols (where the time complexity of the hash function is $T_h$).
\begin{table}[!t]
\begin{center} 
\caption{Comparison of computation cost between authorization protocols} 
\label{tab:performance_comparison}
\scriptsize
\setlength{\tabcolsep}{2pt}
\begin{tabular}{|p{45pt}|p{44pt}|p{40pt}|p{16pt}|p{37pt}|p{32pt}|p{56pt}|p{30pt}|p{21pt}|}
\hline
Protocol                                  &Hashes on $Sen_i$                         &Hashes on $BS$                           &Total                                    &Hash class   &Encryption type       &Execution time in ms    &Requests No.     &Bits No. \\
\hline
\citet{cp92}                            &$1T_{h}$                                          &$2T_{h}$                                       &$3T_{h}$                            &Standard       &Symmetric                   &0.0207                          &6                            &1344\\
\citet{cp91}                            &$4T_{h}$                                          &$9T_{h}$                                       &$13T_{h}$                            &Standard       &-                                    &10.8                            &6                            &704\\
\citet{fp49}                            &$3T_{h}$                                          &$7T_{h}$                                       &$10T_{h}$                            &Standard       &-                                    &0.0001+0.442              &4                            &768\\
\textbf{Proposed}                 &$1T_{h}$                                          &$2T_{h}$                                       &$3T_{h}$                              &Lightweight   &AES-192                   &0.002358                       &6                             &253\\
\hline
\end{tabular}
\end{center}
\end{table}
\section{Conclusion}
\label{sec:conclustion}
One fundamental need for sensor health systems is the security of sensor data.  The accuracy and safety of patients' lives depend heavily on the safeguarding of this sensitive data.  Therefore, we suggested a privacy protocol that was primarily based on Shamir, ECDSA, SAML, and AES-192.  We have demonstrated through the evaluation results that our protocol can prevent attacks on this search field and provides superior performance over the current search performance.  Future trends include expanding security (to include additional assaults), performance testing with more performance metrics, and apps in various health settings.
%\section{Acknowledgements}
%Acknowledgements and Reference heading should be left justified, bold, with the first letter capitalized but have no numbers. Text below continues as normal.

\bibliographystyle{IEEEtranN}
\bibliography{ref}
\end{document}